\documentclass[aps,prf,reprint,onecolumn,groupedaddress,nofootinbib]{revtex4-1}
\usepackage[%
colorlinks=true,
urlcolor=blue,
linkcolor=blue,
citecolor=blue
]{hyperref}

\usepackage{amsmath,esint}
\usepackage{amsbsy}
\usepackage{amsfonts}
\usepackage{graphicx}
\usepackage{epstopdf}
\usepackage{longtable}
\usepackage{scalerel}
\usepackage{stackengine,wasysym}
\usepackage{comment}
\usepackage{placeins}
\usepackage{xcolor}
\usepackage{multirow}
\usepackage[colorinlistoftodos]{todonotes}
\usepackage{psfrag}
\usepackage{scalerel,stackengine}

\stackMath
\newcommand\reallywidehat[1]{
        \savestack{\tmpbox}{\stretchto{
                        \scaleto{
                                \scalerel*[\widthof{\ensuremath{#1}}]{\kern-.6pt\bigwedge\kern-.6pt}
                                {\rule[-\textheight/2]{1ex}{\textheight}}
                        }{\textheight}
                }{0.5ex}}
        \stackon[1pt]{#1}{\tmpbox}
}
\usepackage{tabularx}
\usepackage{enumitem}

\usepackage{graphicx}
\usepackage{natbib}
\usepackage{subfigure}
\usepackage{epstopdf}

\graphicspath{{./eps/}}

\begin{document}

\title{Turbulent burning velocity and thermo-diffusive instability of premixed flames}
        \author{Hsu Chew Lee$^{1,2}$}
        \author{B. Wu$^1$}
        \author{Peng Dai$^1$}
        \author{Minping Wan$^{1,2,3}$}
        \email{wanmp@sustech.edu.cn}
        \author{Andrei N. Lipatnikov$^{4}$}
         \email{andrei.lipatnikov@chalmers.se}
        \affiliation{$^1$Guangdong Provincial Key Laboratory of Turbulence Research and Applications, Department of Mechanics and Aerospace Engineering, Southern University of Science and Technology, Shenzhen 518055, China}
        \affiliation{$^2$Guangdong-Hong Kong-Macao Joint Laboratory for Data-Driven Fluid Mechanics and Engineering Applications, Southern University of Science and Technology, Shenzhen 518055, China}
        \affiliation{$^3$Jiaxing Research Institute, Southern University of Science and Technology, Jiaxing, 314031, Zhejiang, China}
        \affiliation{$^4$Dept. of Mechanics and Maritime Sciences, Chalmers University of Technology, G\"oteborg, 412 96, Sweden}

        \date{\today}

\begin{abstract}
Reported in the paper are results of
unsteady three-dimensional direct numerical simulations of laminar and turbulent, lean hydrogen-air, complex-chemistry flames propagating 
in forced turbulence in a box. 
To explore the eventual influence of thermo-diffusive instability of laminar flames on turbulent burning velocity,
(i) a critical length scale $\Lambda_{n}$ that bounds regimes of unstable and stable laminar combustion is numerically determined by gradually 
decreasing the width $\Lambda$ of computational domain until a stable laminar flame is obtained and
(ii) simulations of turbulent flames are performed by varying the width from $\Lambda<\Lambda_{n}$ (in this case, the instability is suppressed)
to $\Lambda>\Lambda_{n}$ (in this case, the instability may grow).    
Moreover, simulations are performed either using mixture-averaged transport properties (low Lewis number flames) or setting diffusivities of 
all species equal to heat diffusivity of the mixture (equidiffusive flames), with all other things being equal.
Obtained results show a significant increase in turbulent burning velocity $U_T$ when the boundary $\Lambda=\Lambda_{n}$ is crossed in weak 
turbulence, but almost equal values of $U_T$ are computed at $\Lambda<\Lambda_{n}$ and $\Lambda>\Lambda_{n}$ in moderately turbulent flames 
characterized by Karlovitz number equal to 3.4 or larger.
These results imply that thermo-diffusive instability of laminar premixed flames substantially affects burning velocity in weak turbulence only,
in line with a simple criterion proposed by Chomiak and Lipatnikov (Phys. Rev. E 107, 015102, 2023).
\end{abstract}


\maketitle

\section{Introduction}\label{SI}

Since premixed turbulent combustion is a multiscale and highly nonlinear phenomenon \cite{Pet32}, it involves a variety of local effects 
reviewed elsewhere \cite{Dr08,CRC,ARFM,DCSCHW20,SHZ21}, with these effects significantly enriching the physics of turbulent reacting flows.
In particular, rate of premixed combustion is known to be increased not only by a wide spectrum of turbulent velocity fluctuations 
\cite{FAW,KuSa,JC,Pet}, but also by hydrodynamic (Darrieus-Landau, DL) and thermo-diffusive (TD) instabilities of laminar 
flames\footnote{While hydrodynamic instability can arise in any premixed flame characterized by a finite density drop from the unburned side 
to the burned side, thermo-diffusive instability can only arise in mixtures characterized by a low Lewis number $Le<1$ \cite{ZBLM}, 
which is equal to a ratio of molecular mass diffusivity of deficient reactant to molecular heat diffusivity of the mixture. 
Accordingly, a premixed laminar flame characterized by a low $Le$ is subject to both instabilities. 
Henceforth, the term ``DL instability" refers to hydrodynamic instability of a laminar flame characterized by $Le \ge 1$, whereas 
the term ``TD instability" refers to joint effects of both hydrodynamic and thermo-diffusive instabilities on a laminar flame characterized 
by a low $Le < 1$, as the latter effect is stronger.} 
\cite{ZBLM,LL,Cl85,KH05,Mat07,H39}, with the interactions between turbulence and flame instabilities still challenging the combustion community. 
This fundamental challenge has also been attracting rapidly growing interest in applied Computational Fluid Dynamics (CFD) research aiming at  
developing zero-emission technologies for power generation by utilizing chemical energy bound in renewable carbon-free fuels such as hydrogen.
The point is that 
(i) since pioneering experiments by Karpov et al. \cite{KS61,KS80},
significant differences between molecular diffusivities of H$_2$, O$_2$, and heat were well known to result in  
strongly increasing turbulent burning velocities $U_T$ in sufficiently lean hydrogen-air mixtures, as reviewed elsewhere \cite{KuSa,H39,PECS05},
see also recent experimental studies \cite{YSLWL18,NYS37,LCS39}, 
(ii) a similar phenomenon was recently documented in experiments with fuel blends that contain H$_2$, 
e.g., lean syngas/air mixtures \cite{VMSNSL11,DJMB33,VMSL34,ZWYJZH18} or lean ammonia/hydrogen/air mixtures 
\cite{XHHHHKF20,LBCMR38,WEAR23,WEWWR23,CFBWZHAL39},
(iii) a widely established predictive model of this important effect has yet to be developed, while 
(iv) the effect is often discussed \cite{H39,NYS37,XHHHHKF20,WEAR23,ADB11,CC11,KOWMOOKK34,DCC17,ZWH20,BAP22b} in terms of 
TD instability of laminar premixed flames.   

Indeed, on the one hand, both the increase in $U_T$ and TD instability result from variations in local burning rate, 
caused by imbalance of molecular fluxes of chemical and thermal energies to and from, respectively, reactions zones stretched by the local 
velocity field \cite{ZBLM,KuSa,CRC,BLL92}.
This similarity of the governing physical mechanisms encourages researchers to link the two phenomena. 
However, 
on the other hand, the present authors are not aware of a convincing evidence that the increase in $U_T$ stems from local instabilities 
of inherently laminar flames in a turbulent flow or that the increase magnitude can be predicted by invoking characteristics of TD instability, 
such as its growth rate.

On the contrary, the following simple order-of-magnitude criterion 
\begin{equation}
\label{EKacr}
Ka = \frac{\tau_f}{\tau_K} < Ka_{TD}^{cr} = \sqrt{15} \tau_f \max\{\omega_{TD}(k)\}
\end{equation}
of importance of TD instability in turbulent flows has recently been introduced \cite{CL23} by highlighting mitigation of flame instabilities 
by normal flame strain rates \cite{BL82,SLJ82,KBS97}, which are created by small-scale eddies in turbulent flows. 
Here, $Ka$ is a Karlovitz number; 
$\tau_f=\delta_L/S_L$, $S_L$, and $\delta_L=(T_b-T_u)/\max\{|dT/dx|\}$ are the laminar flame time scale, speed, and thickness, respectively; 
$T$ is the temperature;
subscripts $u$ and $b$ designate unburned reactants and burned products;
$\max\{\omega_{TD}(k)\}$ is the maximum growth rate of TD instability (dependence of the instability growth rate $\omega_{TD}$ on the 
instability wavenumber $k$ has a bell-like shape \cite{AFTKB33,AFTMB12,FFTAM35,BKAP19,BAP22a,BGJLCAP39}, as discussed in the next section); and 
$\tau_K$ is Kolmogorov time scale, which characterizes the order of magnitude of the highest velocity gradients generated by the smallest-scale 
turbulent eddies \cite{LL,MY,Frisch}.
Similar criteria of importance of DL instability in turbulent flows was earlier introduced using the same reasoning \cite{PECS05}, 
as well as other reasoning \cite{BT27,CAL11}, and results of subsequent numerical \cite{CM11,FCM34,FCM15,FCM17} and experimental \cite{LTLC38}  
studies qualitatively supported those criteria by indicating a minor role played by DL instability under conditions of sufficiently intense 
turbulence. 
However, to the best of the present authors knowledge, the criterion given by Eq. (\ref{EKacr}) has been neither supported nor disputed 
in studies of thermo-diffusively unstable flames and target-directed research into eventual mitigation of TD instability effects in 
turbulent flows is yet to be done.

Thus, a role played by TD instability of laminar premixed flames in turbulent flows is still an intricate fundamental and 
practically important issue. 
The present paper aims at clarifying it by reporting results of a target-directed Direct Numerical Simulation (DNS) study. 
In the next section, adopted research method is presented and the DNS attributes are summarized.
Obtained results are discussed in section III, followed by conclusions.

\section{Method} \addvspace{10pt} 
\label{SM}

\subsection{Key point} \addvspace{10pt}
\label{sSKP} 

The present work is based on a seminal idea put forward and developed recently by Matalon et al. \citep{CM11,FCM34,FCM15,FCM17} who
numerically studied interactions of DL instability with turbulence. 
This idea stems from the following considerations.
Both theoretical \citep{PC82,MM82,FS82,CMK03,KBL12} and experimental \citep{CS98,TS99} studies of DL instability show that dependence of 
the instability growth rate $\omega_{DL}$ on perturbation wavenumber $k$ has a bell-shape form, i.e., a peak growth rate $\omega_{DL}>0$ is 
reached at certain wavenumber $k_m$, whereas molecular transport processes stabilize the flame with respect to small-scale perturbations 
whose wavenumber is larger than the neutral wavenumber $k_n$, with $k_n \approx 2 k_m$ and $\omega_{DL}(k>k_n)<0$.
Accordingly, variations in a computational domain width $\Lambda$ offer the opportunity to numerically explore
a role played by DL instability in premixed turbulent combustion by comparing results simulated under conditions of $\Lambda < 2 \pi/k_n$ 
(the instability is suppressed) and $\Lambda > 2 \pi/k_n$ (the instability can grow), with all other things being equal.
Results obtained in such simulations \citep{CM11,FCM34,FCM15,FCM17} showed that DL instability could substantially contribute to $U_T$ 
under conditions of weak turbulence only, in line with simple phenomenological criteria proposed earlier \citep{PECS05,BT27,CAL11}.

To the best of the authors' knowledge, this elegant idea has not yet been adapted in research into TD instability, probably, 
because neither measured nor theoretical (in the case of substantial density variations) dispersion relation $\omega_{TD}(k)$ is known. 
However, recent two-dimensional numerical simulations of laminar premixed flames subject to TD instability 
\citep{AFTKB33,AFTMB12,FFTAM35,BKAP19,BAP22a,BGJLCAP39} have partially  
bridged this knowledge gap by showing that $\omega_{TD}(k)$ has a bell-shape form also, with 
$\omega_{TD}(k=k_m)=\max\{\omega_{TD}(k)\}>0$, $\omega_{TD}(k>k_n)<0$, and $k_n \approx 2 k_m$.
Therefore, the influence of TD instability on turbulent burning velocity may be explored by comparing DNS results 
obtained in the cases of $\Lambda < \Lambda_n = 2 \pi/k_n$ (the instability is suppressed) and $\Lambda > \Lambda_n$ (the instability can grow), 
with all other things being equal.

This approach is implemented in the present work by simulating lean (the equivalence ratio $\phi=0.5$) complex-chemistry hydrogen-air flames
propagating under room conditions either in a laminar flow or in forced turbulence in a box. 
It is worth stressing, however, that the primary goal of the present study calls for specific conditions of the performed DNSs.

First, since (i) the study aims at comparing results computed in domains whose widths are comparable with 
(slightly smaller and slightly larger than) $\Lambda_n$ and (ii) $\Lambda_n$ is sufficiently small for the adopted mixture, see section II.C, 
the DNSs should be run in narrow computational domains, narrower than computational domains used in a typical DNS study of 
a complex-chemistry turbulent premixed flame.
Accordingly, the present work does not aim at exploring evolution of TD instability in a laminar or turbulent flow, 
because much wider computational domains are known to be required to numerically predict major characteristics of a non-linear stage of 
the growth of a laminar flame instability \cite{YBB15,CLLFM20,BAP22c}.
As the present work aims solely at investigating eventual suppression of TD instability by turbulence,
the focus of the study is solely placed on the onset (if any) of the instability in various turbulent flows.       

Second, as discussed in section II.C, Eq. (\ref{EKacr}) assumes that, for the studied mixture, TD instability is suppressed by moderately intense 
turbulence characterized by $Ka={\mathrm O}(1)$, while $Ka>1$.
Since the restriction on the computational domain width, highlighted above, requires considering small-scale turbulence (its integral length 
scale $L$ should be significantly less than the width $\Lambda$) and $Ka \propto (u'/S_L)^{3/2} (L/\delta_L)^{-1/2}$,  
the critical value $Ka_{TD}^{cr}$ yielded by Eq. (\ref{EKacr}) can be reached at sufficiently low rms turbulent velocities $u'$.
Such a weak and small-scale turbulence is characterized by a sufficiently low Reynolds number and even the use of the word ``turbulence" 
may be disputed, e.g., due to the lack of the inertial range of turbulence spectrum. 
Nevertheless, such unusual conditions may still be appropriate for the major goal of the present study. 
Indeed, if so weak turbulence can suppress TD instability, as will be shown in section III,  
there is no reason to expect that the instability could arise in more intense turbulence characterized by a higher $Ka$. 
If TD instability can play a role in weakly and moderately turbulent flames only, running DNS of highly turbulent flames to explore TD 
instability effects does not seem to be a worthy task.

Thus, when the present DNS study was started, its conditions were set using a constraint of $\Lambda \approx \Lambda_n$ and 
Eq. (\ref{EKacr}).
Since the computed results agreed with Eq. (1), there was no need for running DNS by increasing $u'/S_L$,
especially as DNS results obtained recently from highly turbulent flames propagating in wider boxes were already reported by us
\cite{JFM21,CNF22PI,CNF22PII,PoF22,IJHE22,Fuel22}.

\subsection{DNS attributes} \addvspace{10pt}
\label{sSDNSa}

Since the present DNSs are basically similar to our simulations discussed in detail earlier
\cite{JFM21,CNF22PI,CNF22PII,PoF22,IJHE22,Fuel22}, only a summary of the DNS attributes is given below.

Unsteady three-dimensional simulations of statistically one-dimensional and planar 
flames propagating under room conditions in forced turbulence (or a laminar flow) in a box were performed using 
a detailed chemical mechanism (9 species and 22 reversible reactions) by \citet{Curran},
with mixture-averaged molecular transport and chemical reaction rates being modeled using open-source library Cantera-2.3 \citep{Cantera}.
Navier-Stokes, energy, and species transport equations written in the low-Mach-number formulation were numerically integrated using
solver DINO \citep{DINO}. 
It adopts a 6th-order finite-difference central stencil and a semi-implicit 3rd-order Runge–Kutta method for time advancement. 

A rectangular computational domain of $16 \Lambda \times \Lambda \times \Lambda$  
was discretized using a uniform Cartesian grid of $16 N \times N \times N$ cells.
The adopted numerical meshes ensured more than 20 grid points across the thickness $\delta_L$ in the majority of the studied cases,
while the number of grid points per $\delta_L$ was less (15) in two equidiffusive flames discussed later.
In all cases, half the Kolmogorov length scale $\eta_K$ was larger than the grid size $\Delta x$, see table \ref{T1}.
Along the streamwise direction $x$, inflow and outflow boundary conditions were set.
Other boundary conditions were periodic.

Within a rectangular domain of $0.5 \Lambda \le x \le 8 \Lambda$, turbulence was generated using the linear velocity forcing method 
\citep{Lu03,RM05,CB14} and the evolution of this turbulence was simulated for at least 50 integral time scales
$\tau_t=L/u'$ before embedding the steady planar laminar flame solution obtained using Cantera-2.3 \citep{Cantera} into the 
computational domain.
Here, $u'=\overline{\langle u'_i u'_i\rangle}/3$ is rms turbulent velocity; 
the integral length scale $L={u'_0}^3/\overline{\langle \varepsilon \rangle}_0$ yielded by the adopted 
forcing method is about $0.19 \Lambda$ \citep{Lu03,RM05,CB14}; 
$\varepsilon=2 \nu S_{ij} S_{ij}$ is the dissipation rate of turbulent kinetic energy;
$\nu$ is the kinematic viscosity of the mixture;
$S_{ij}=(\partial u_i/\partial x_j + \partial u_j/\partial x_i)/2$ is rate-of-strain tensor;
$u_i$ is $i$-th components of velocity vector;
overline and angle brackets refer to time and transverse-averaged quantities, respectively;
summation convention applies to repeated indexes ($i$ or $j$);
and subscript 0 refers to the constant-density non-reacting turbulent flow simulated before embedding the flame into the computational domain.
In turbulent flame brush, $u'$ varies weakly, while $\overline{\langle \varepsilon \rangle}$ increases gradually along the axial direction 
\cite{CNF22PI}.
The combustion simulations were run for at least $30 \tau_t$.

\subsection{DNS conditions} \addvspace{10pt}
\label{sSDNSc}

\begin{table}
  \begin{center}
  \caption{Characteristics of DNS cases.}
  \label{T1}
  \begin{tabular}{ccccccccccccc}
  \hline
  \hline
  Case  & $S_L$, &  $\delta_L$,  & $u'/S_L$  &  $L/\delta_L$   & $Re_{\lambda}$ & $Ka$ & $Da$ & $\Delta x/L$ & $\Delta x/\eta_K$
&  $N$  & $\Lambda$ & $\overline{U_T}/S_L$ \\
        &  m/s   &       mm      &           &                  &                  &      &      &              &
&       &     mm    &                      \\
  \hline
  LT/S  &  0.58  &      0.41     &    0.34   &      0.58        & 7.9              & 0.9  & 1.8  &    0.08      &      0.21
&  64   &    1.26   &          1.05        \\
  LT/U  &  0.58  &      0.41     &    0.34   &      0.61        & 7.5              & 0.9  & 1.8  &    0.08      &      0.22
&  64   &    1.32   &          1.64        \\
  WT/U  &  0.58  &      0.41     &    0.50   &      0.61        & 9.8              & 1.6  & 1.2  &    0.08      &      0.30
&  64   &    1.32   &          1.71        \\
  T/S   &  0.58  &      0.41     &     1.0   &      0.58        & 13.7             & 4.4  & 0.6  &    0.04      &      0.24     
&  128  &    1.26   &          1.90        \\
  T/U   &  0.58  &      0.41     &     1.0   &      0.61        & 14.0             & 4.3  & 0.6  &    0.04      &      0.25
&  128  &    1.32   &          1.93        \\
  T/UM  &  0.58  &      0.41     &     1.0   &      1.1         & 18.0             & 3.4  & 1.1  &    0.04      &      0.40
&  128  &    2.4    &          3.29       \\ 
  T/U1  &  0.78  &      0.29     &    0.74   &      0.61        & 14.7             & 2.2  & 1.2  &    0.04      &      0.24
&  128  &   1.32    &          1.10       \\
  T/UM1 &  0.78  &      0.29     &    0.74   &      1.1         & 18.0             & 1.8  & 2.1  &    0.04      &      0.40
&  128  &    2.4    &          1.73        \\
  \hline
  \hline
  \end{tabular}
  \end{center}
\end{table}

The simulation conditions are summarized in table \ref{T1}, where 
$Re_{\lambda}=u' \lambda/\nu$ is the turbulent Reynolds number based on the Taylor length scale 
$\lambda=u'(15 \nu/\overline{\langle \varepsilon \rangle})^{1/2}$;
$Ka=\tau_f/\tau_K$, and $Da=\tau_t/\tau_f$ are turbulent Karlovitz and Damk\"ohler numbers, respectively;
$\eta_K=(\nu^3/\overline{\langle \varepsilon \rangle})^{1/4}$ and $\tau_K=(\nu/\overline{\langle \varepsilon \rangle})^{1/2}$ are 
Kolmogorov length and time scales, respectively, with
the time and transverse-averaged dissipation rate $\overline{\langle \varepsilon \rangle}$ being averaged over flame-brush 
leading edge characterized by $\langle c_F \rangle(x,t)=0.01$;
$c_F=1-Y_F/Y_{F,u}$ designates combustion progress variable evaluated using the fuel mass fraction $Y_F$;
and $\Delta x=\Lambda/N$ is the grid size.
Reported in the last column are normalized time-averaged values of turbulent burning velocities,     
which have been evaluated by integrating the fuel consumption rate
$\dot{\omega}_{H_2} (\boldsymbol{x},t)$ over the computational domain, i.e.,
\begin{equation}
\label{EUt} 
U_T(t) = \frac{1}{(\rho Y_{H_2})_u \Lambda^2} \int\!\!\int\!\!\int \left | \dot{\omega}_{H_2} \right | (\boldsymbol{x},t) d \boldsymbol{x}.
\end{equation}
Here, $\rho$ is the density.

When compared to our earlier simulations \citep{JFM21,CNF22PI}, three differences should be emphasized.
First, the rms velocity $u'$ was reduced, because TD instability was assumed to be suppressed in highly turbulent flames and this assumption 
was supported in subsequent simulations, as discussed in section III.
Accordingly, letters LT, WT, and T in case names in table \ref{T1} refer to transition from laminar to turbulent flows, weak turbulence, 
and moderately turbulent flames, respectively.

\begin{figure}
\centerline{
\subfigure[]{
\label{Flfdr}
\includegraphics[height=5.2cm,width=6.3cm]{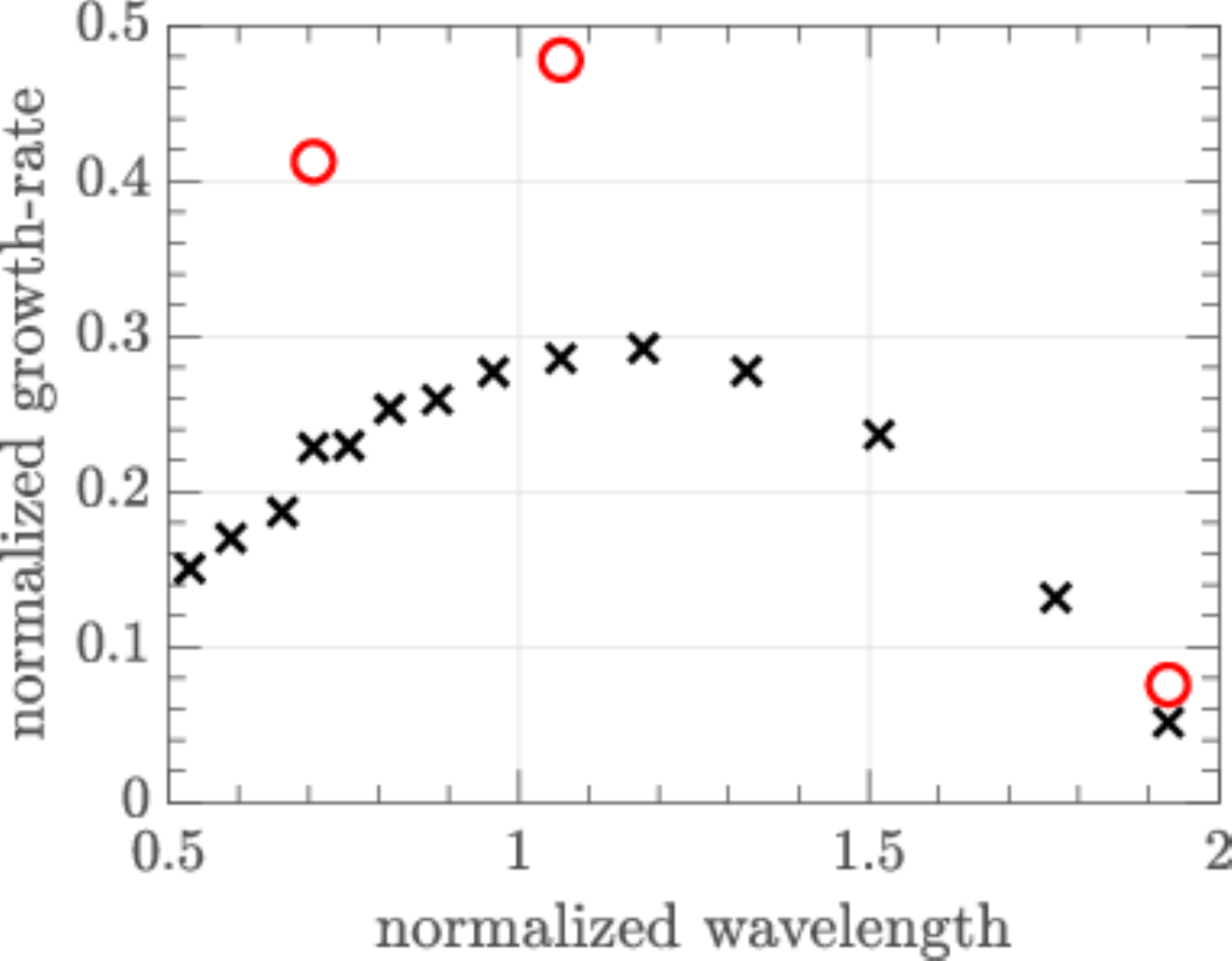} 
}%
\subfigure[]{
\label{FlfUS}
\includegraphics[height=5.2cm,width=6.3cm]{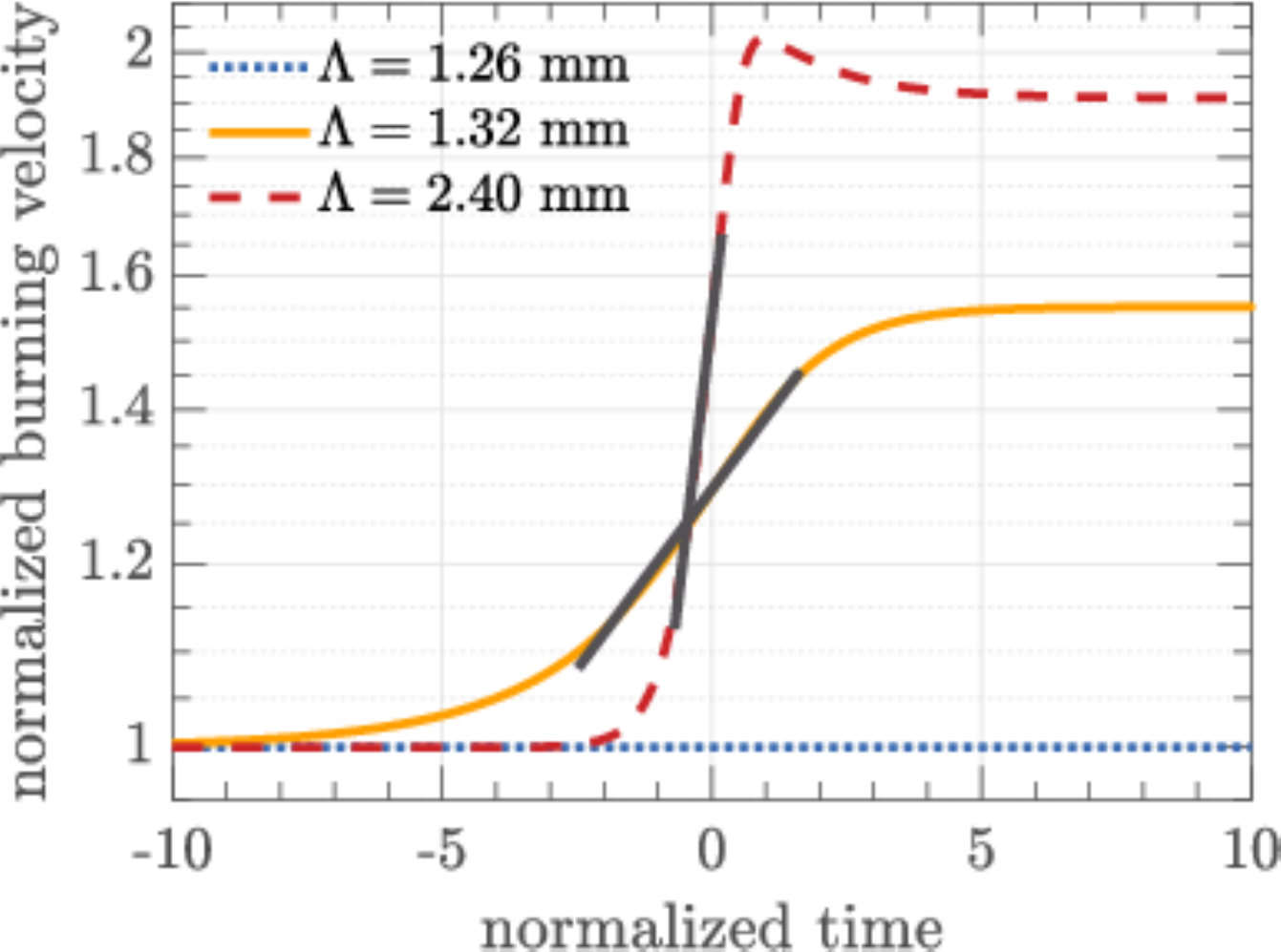}
}
}
\caption{
\label{Flf}
(\textit{a}) Normalized instability growth rates $\tau_f \omega_{TD}(k)$ obtained from two-dimensional (black crosses) and three-dimensional 
(red circles) laminar flames by varying the width $\Lambda$ of the computational domain and the perturbation
wavelength $k=2 \pi/\Lambda$, with all other things being equal.
(\textit{b}) Normalized laminar burning velocities computed by varying $\Lambda$ and $k=2 \pi/\Lambda$.
In two unstable cases, the normalized time $t/\tau_f$ is shifted to get the maximum rate of an increase in the burning velocity at $t'=0$.
To evaluate the instability growth rate, the ordinate axis is scaled in the natural-logarithm units and
thick black straight lines fit the computed curves. 
}   
\end{figure} 

Second, the domain width $\Lambda$ was varied from $\Lambda < \Lambda_n$ to $\Lambda > \Lambda_n$, with the neutral width $\Lambda_n$
being found in pre-simulations of two-dimensional and three-dimensional laminar flames. 
In the two-dimensional case, $\Lambda$ was changed with a small step and
weak periodic velocity perturbations with the wavenumber $k=2 \pi/\Lambda$ were generated at the inlet.
Those pre-simulations were run adopting the same chemical and molecular transport models, the same solver, and                                 
$N=128$, with the independence of the numerical results on the spatial resolution being checked using $N=256$ or 512.     
Subsequently, these results were adopted to set conditions of a few three-dimensional pre-simulations, 
which were performed to find $\Lambda_n$ in the three-dimensional case. 

The results of these pre-simulations, reported in Fig. \ref{Flf}, see also Figs. \ref{FimLFS} and \ref{FimLFU} in the next section, 
show that the laminar flame is stable at $\Lambda=1.26$ mm, but unstable at $\Lambda=1.32$ mm, both in the two-dimensional 
and three-dimensional cases.
For instance, instability growth rates computed in the two-dimensional and three-dimensional cases are plotted in black crosses and red circles, 
respectively, in Fig. \ref{Flfdr}.
As illustrated in Fig. \ref{FlfUS}, these growth rates have been evaluated by (i) computing laminar burning velocities adapting Eq. (\ref{EUt}) 
and (ii) fitting the linear parts of the obtained dependencies of $\ln(U_L/S_L)$ on time with straight lines. 
Dispersion relations $\omega(k)$ similar to the curve plotted in black crosses in Fig. \ref{Flfdr} are well known from 
theoretical \citep{PC82,MM82,FS82,CMK03,KBL12} and experimental \citep{CS98,TS99} studies of DL instability, as well as from 
recent two-dimensional numerical simulations of thermo-diffusively unstable laminar flames 
\citep{AFTKB33,AFTMB12,FFTAM35,BKAP19,BAP22a,BGJLCAP39}.
As far as quasi-stationary values of $\ln(U_L/S_L)$, reached at large $t/\tau_f$, see Fig. \ref{FlfUS}, and associated with a non-linear stage 
of the instability development, are concerned, they are expected to be increased with an increase in the computational domain width $\Lambda$. 
However, this non-linear stage was already investigated in recent two-dimensional numerical simulations \cite{CLLFM20,BAP22c} performed using 
significantly wider computational domains and, therefore, is beyond the scope of the present study, whose focus is solely placed on the onset 
of TD instability. 
 
Based of the presented results of the laminar flame pre-simulations, 
letter S or U in case names in table \ref{T1} refer to $\Lambda=1.26$ mm (TD instability cannot arise, stable case) and $\Lambda=1.32$ mm 
(the instability may occur, unstable case), respectively.
The major goal of the present study is to compare turbulent burning velocities $U_T$ computed in LT/S and LT/U or T/S and T/U cases.

Third, if $\Lambda$ is slightly above $\Lambda_n$, the growth rate $\omega_{TD}(k)$ of allowed unstable perturbations is low,
see Fig. \ref{Flfdr} or compare the maximum slopes of curves plotted in yellow solid and red dashed lines in Fig. \ref{FlfUS}.
Therefore, even if TD instability with respect to such perturbations is of minor importance when compared to turbulence,
perturbations with the maximum growth rate $\omega_{TD}(k_m)$ might significantly affect turbulent burning velocity.
To explore such a scenario, the width $\Lambda$ was increased to 2.4 mm, i.e., by a factor of about two when compared to $\Lambda_n$.
In this case TUM, labeled with extra letter M in table \ref{T1}, appearance of perturbations whose growth rates were close to 
$\max\{\omega_{TD}(k)\}$ was enabled, see the most-up circle in Fig. \ref{Flfdr}.

However, the aforementioned increase in $\Lambda$ can increase $U_T$ not only due to development of TD instability, but also due to an increase in
the scale $L$, because burning velocity is increased by turbulence length scale, as reviewed elsewhere \citep{PECS02}, see also a recent
DNS study by \citet{YL17} or a recent experimental work by \citet{KSLG20}.
To compare magnitudes of these two effects, two more cases T/U1 and T/UM1 (in addition to cases T/U and T/UM, respectively) were run by setting
molecular diffusivities $D$ of all species equal to molecular heat diffusivity $\kappa$ of the mixture.
Number 1 in the names of these two cases shows that the Lewis number $Le=\kappa/D_{H_2}=1$.
In other cases, $Le=0.32$.
Since TD instability does not appear if $Le=1$ \citep{ZBLM}, a ratio $R1$ of turbulent burning velocities computed in cases T/UM1 and T/U1 
characterizes magnitude of the latter effect (an increase in $U_T$ by $L$).
Therefore, comparison of a ratio $R$ of $U_T$ obtained in cases T/UM and T/U with the ratio $R1$ offers the opportunity to estimate importance
of the former effect (TD instability).

\begin{figure}
\centerline{
\subfigure[]{
\label{FUta}
\includegraphics[height=5.2cm,width=6.3cm]{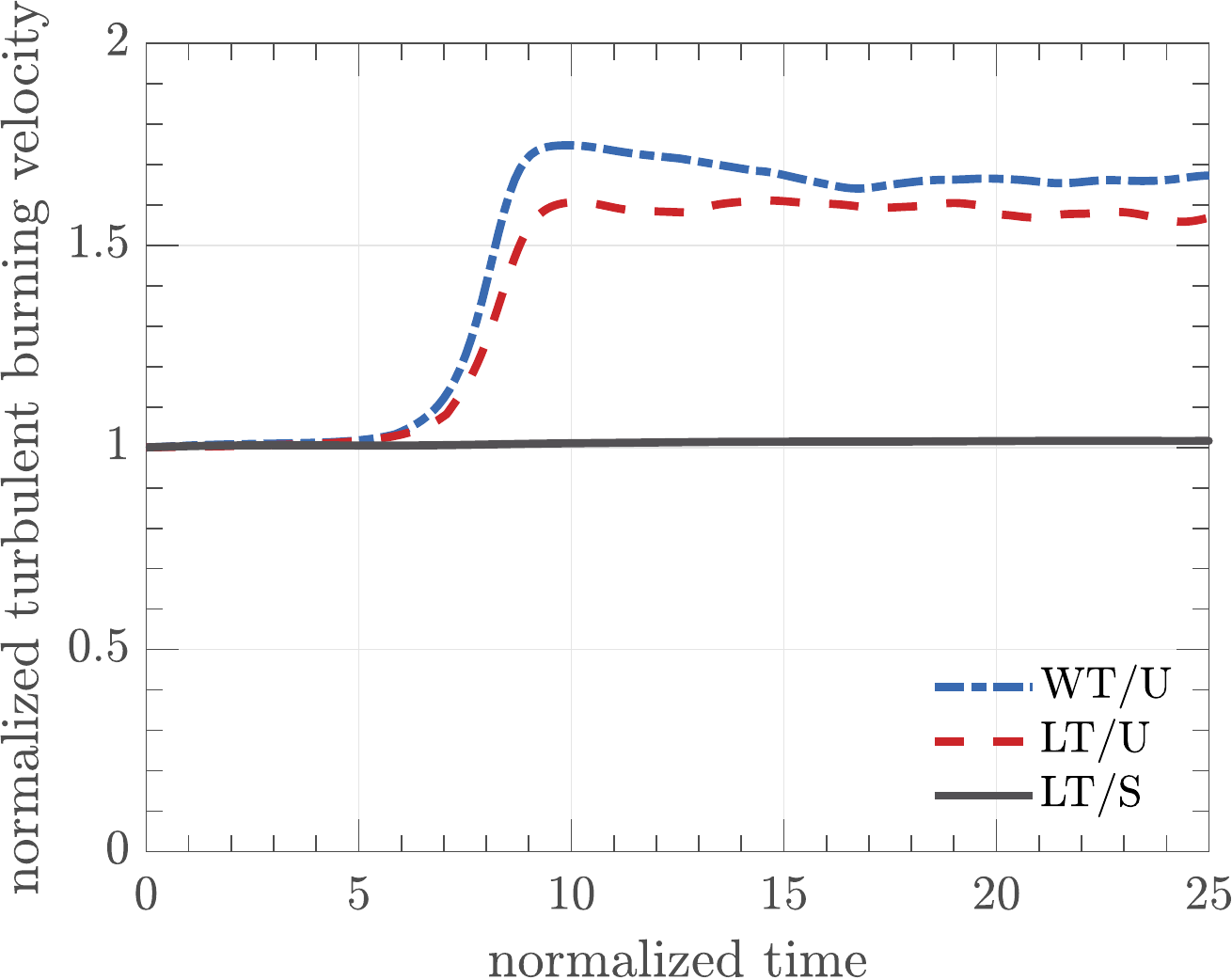}
}
\subfigure[]{
\label{FUtb}
\includegraphics[height=5.2cm,width=6.3cm]{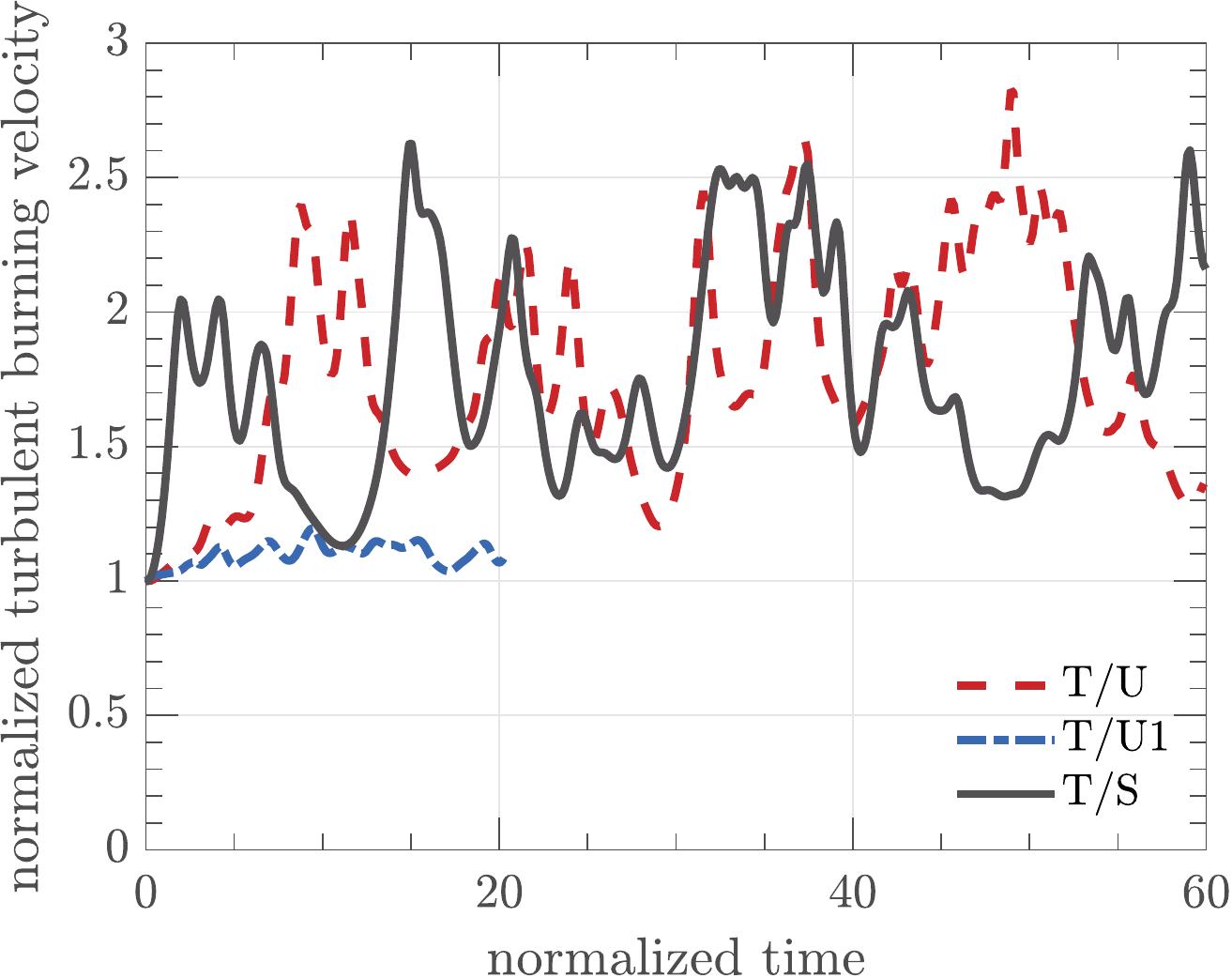}
}
\subfigure[]{
\label{FUtc}
\includegraphics[height=5.2cm,width=6.3cm]{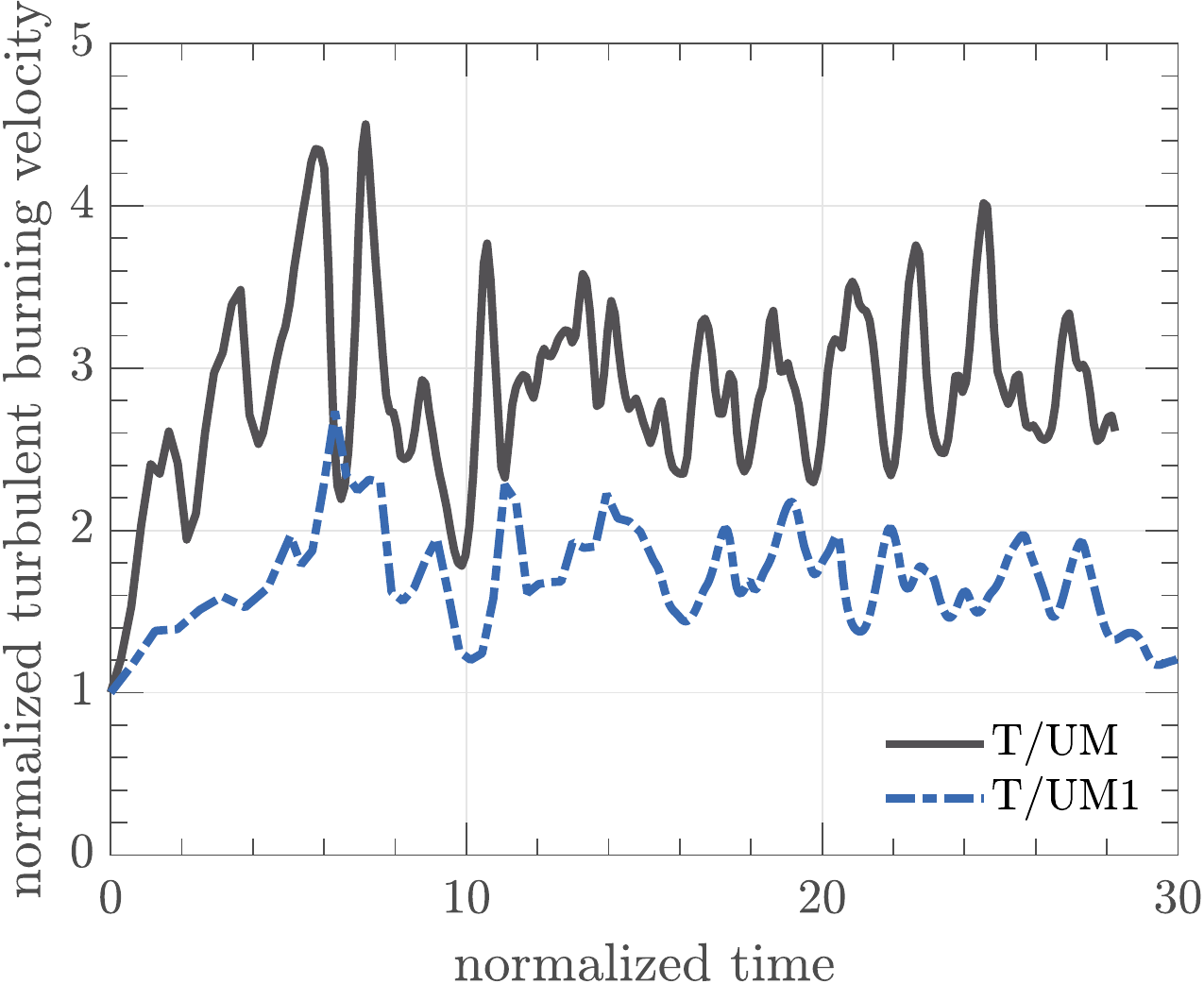}
}
}
\caption{ 
\label{FUt}
Evolution of the normalized turbulent burning velocities $U_T/S_L$ in flames
(\textit{a}) LT/S (black solid line), LT/U (red dashed line), and WT/U (blue dotted-dashed line),
(\textit{b}) T/S (black solid line), T/U (red dashed line), and T/U1 (blue dotted-dashed line).
(\textit{c}) T/UM (black solid line), and T/UM1 (blue dotted-dashed line).
Time is normalized using the laminar flame time scale $\tau_f$.
}
\end{figure}

\section{Results and discussion}\label{SRD}

\begin{figure}
\centerline{ 
\subfigure[]{
\label{FimLFS}
\includegraphics[height=5.2cm,width=6.3cm]{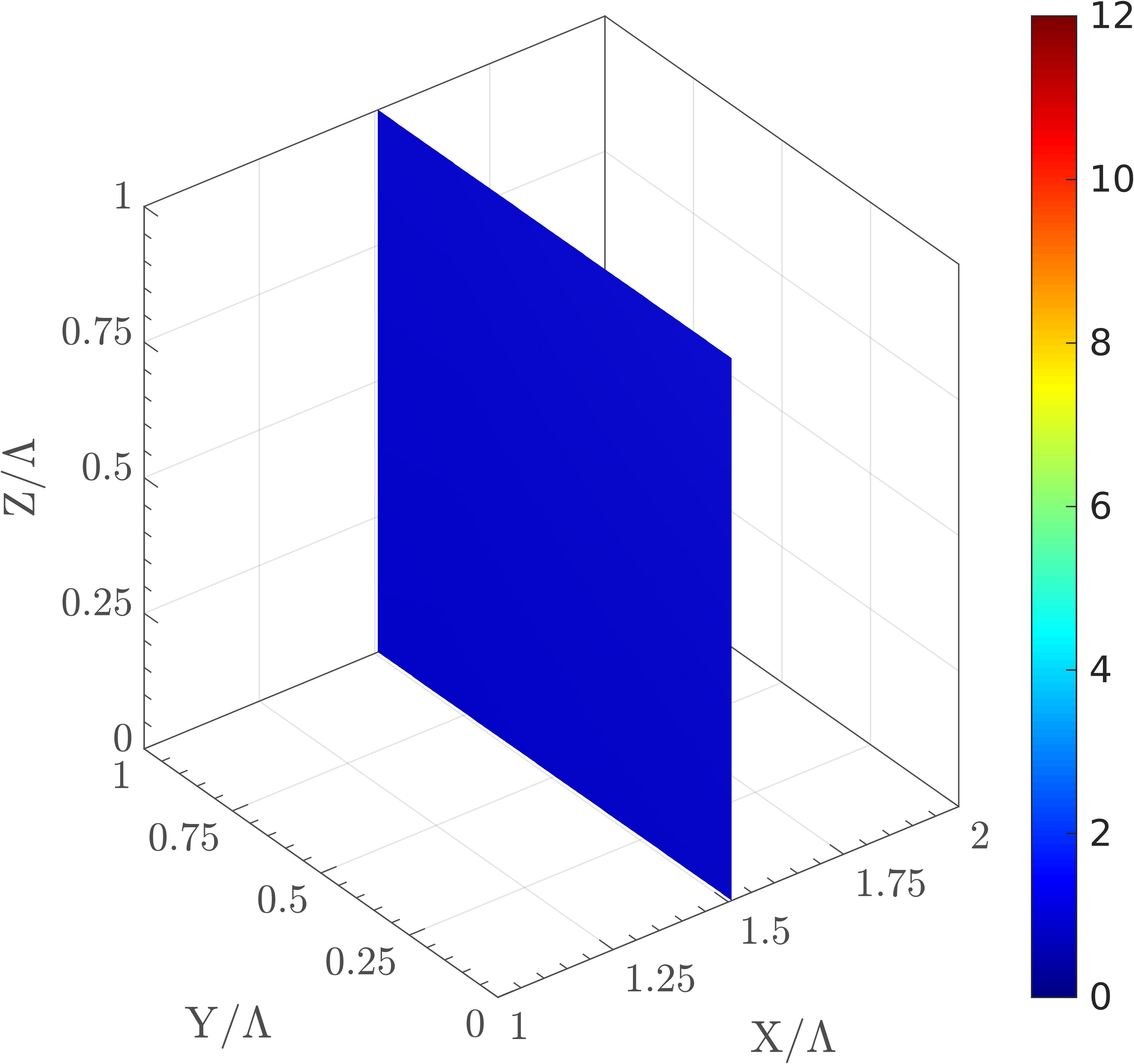}
}%
\subfigure[]{
\label{FimLFU}
\includegraphics[height=5.2cm,width=6.3cm]{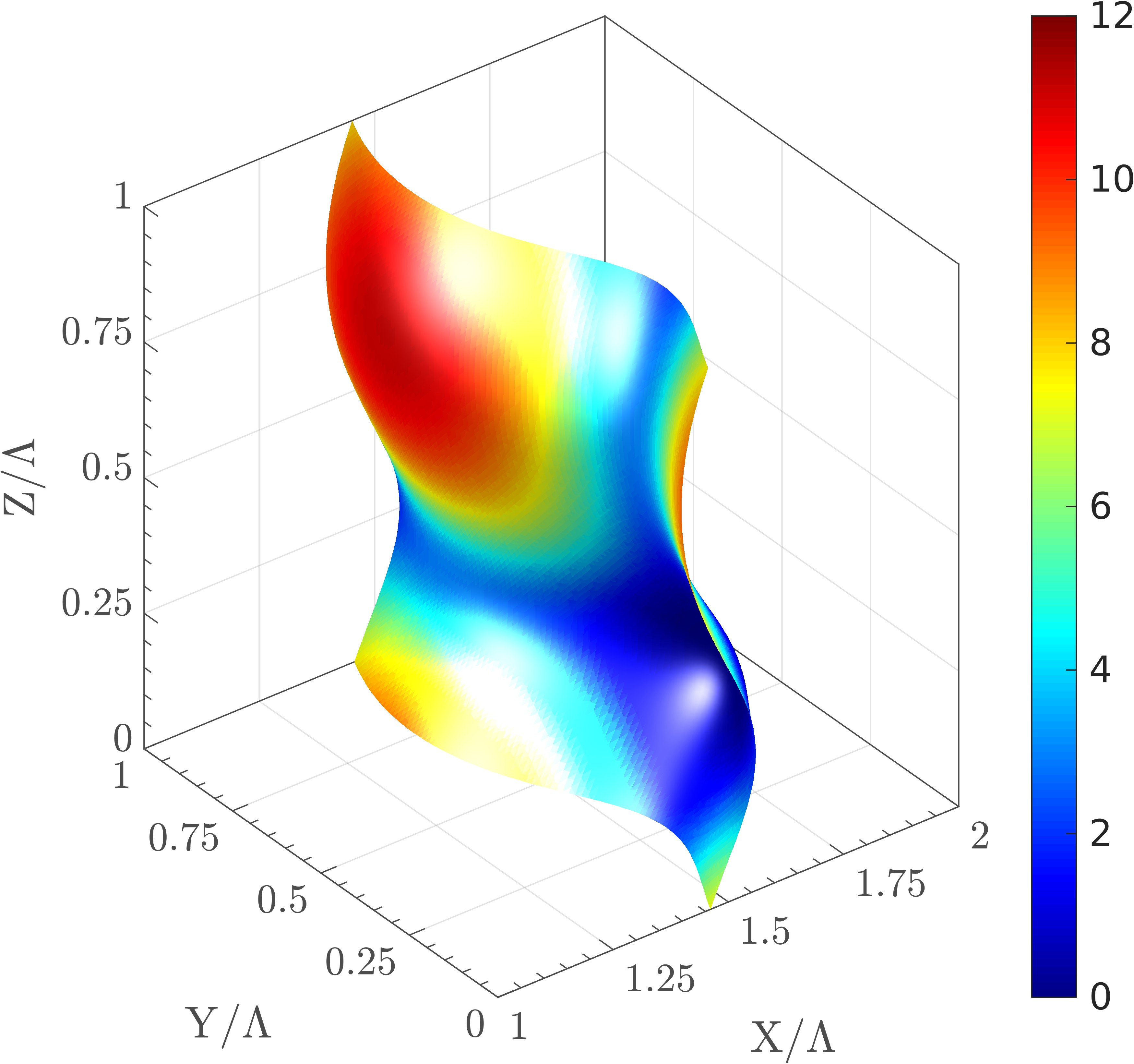}
}
}
\centerline{
\subfigure[]{
\label{FimLTFS} 
\includegraphics[height=5.2cm,width=6.3cm]{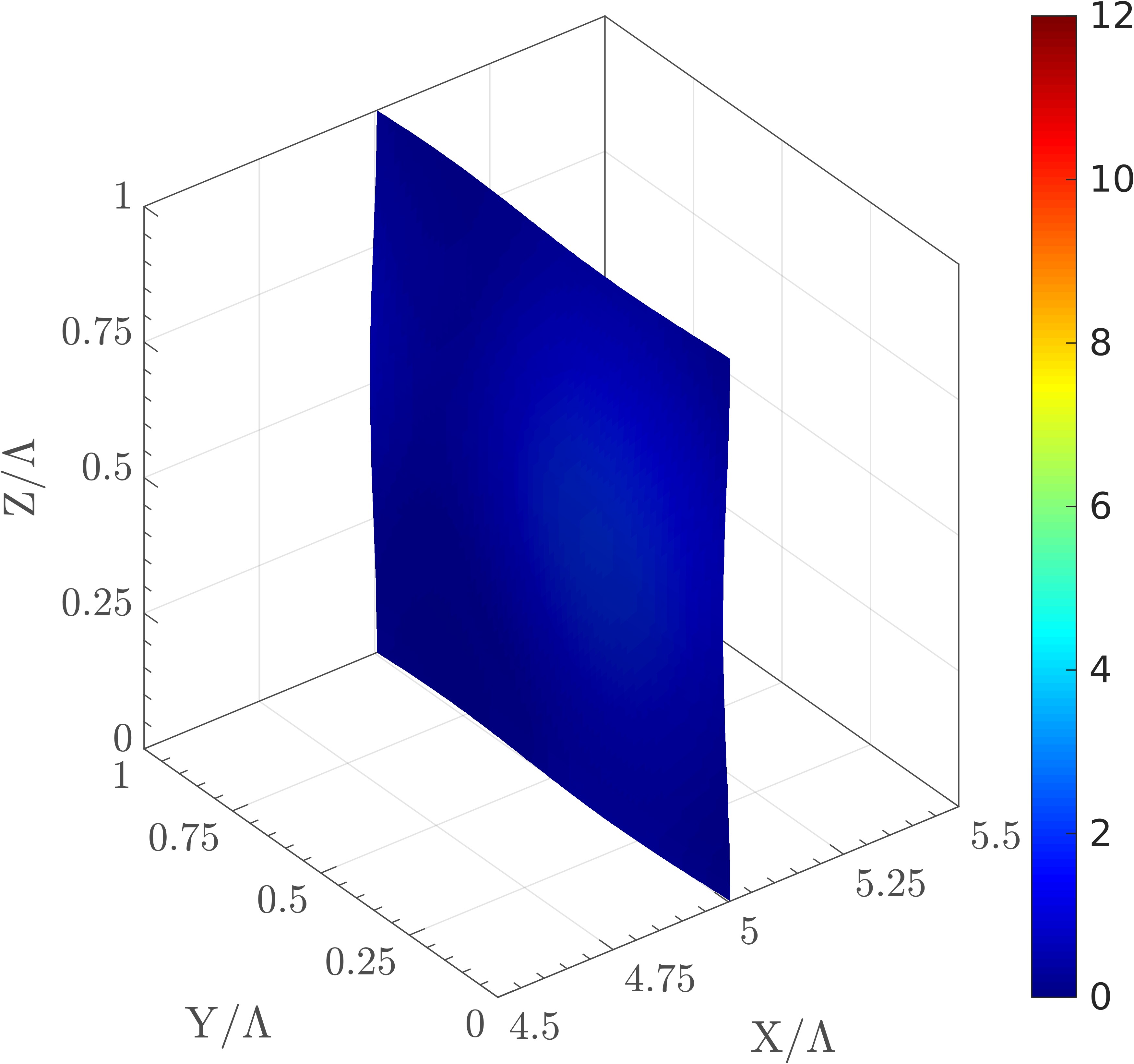}
}
\subfigure[]{ 
\label{FimLTFU}
\includegraphics[height=5.2cm,width=6.3cm]{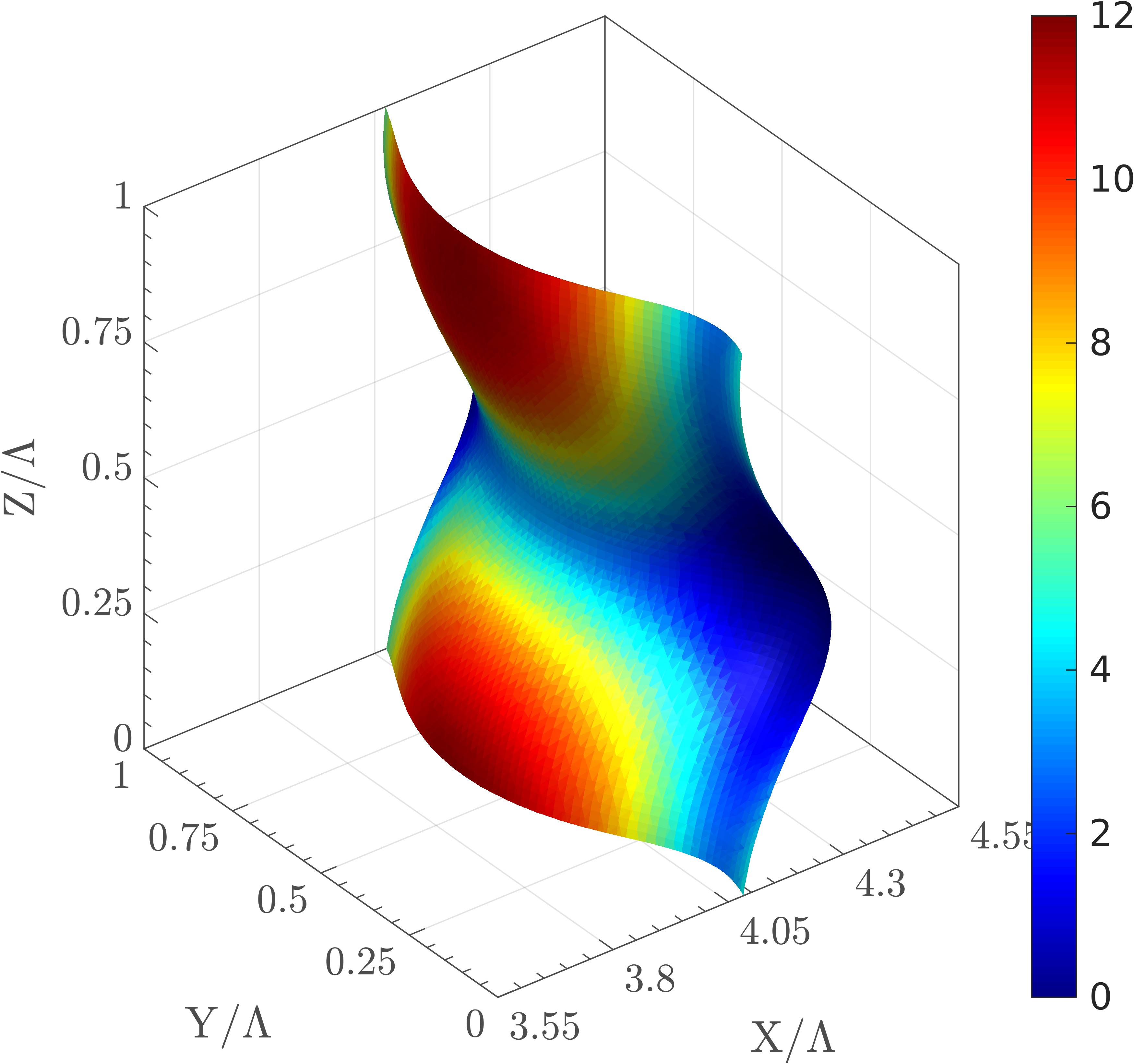}
}
}
\centerline{
\subfigure[]{
\label{FimTFS}
\includegraphics[height=5.2cm,width=6.3cm]{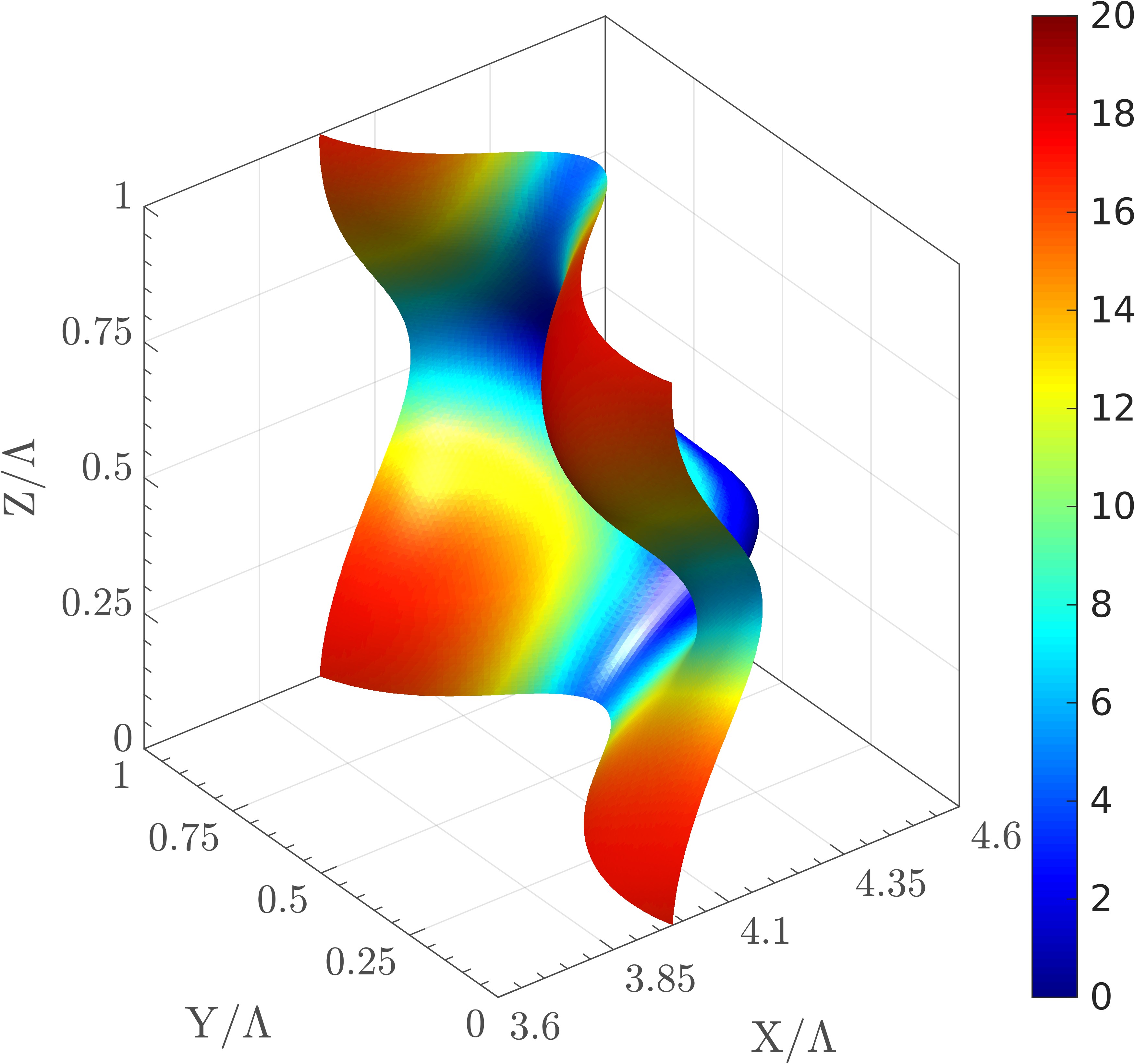}
}
\subfigure[]{
\label{FimTFU}
\includegraphics[height=5.2cm,width=6.3cm]{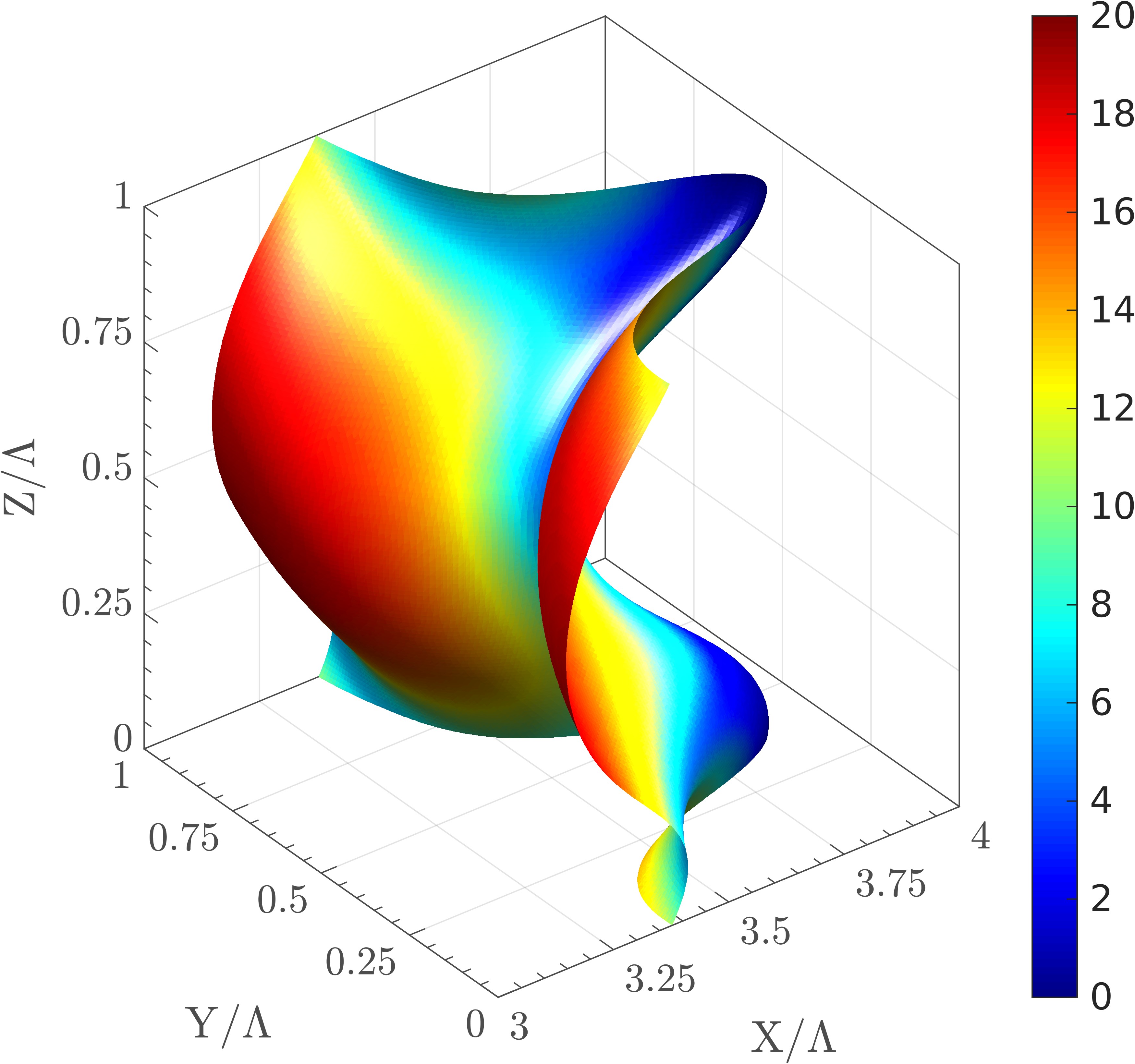}
}
}
\caption{
\label{Fim}
Images of the instantaneous iso-surfaces $c_F(\boldsymbol{x},t)=0.5$, with color bars showing the local fuel consumption rate
$\dot{\omega}_{H_2}(\boldsymbol{x},t)$ normalized with the rate $\dot{\omega}_{H_2,L}(c_F=0.5)$ obtained from the unperturbed laminar flame.
(\textit{a}) stable laminar flame, $\Lambda=1.26$ mm,
(\textit{b}) unstable laminar flame, $\Lambda=1.32$ mm,
(\textit{c}) flame LT/S,
(\textit{d}) flame LT/U,
(\textit{e}) flame T/S, and
(\textit{f}) flame T/U.
}
\end{figure}

Computed dependencies of the normalized turbulent burning velocity $U_T(t)/S_L$ on the normalized time $t/\tau_t$ are reported in Fig. 
\ref{FUt}, with representative images of instantaneous turbulent flame surfaces being shown in Figs. \ref{FimLTFS}-\ref{FimTFU}.
The following trends are worth noting.

First, in case LT/S, flame surface looks weakly perturbed, see Fig. \ref{FimLTFS}, and $U_T(t) \approx S_L$, 
see curve plotted in black solid line in Fig. \ref{FUta}. 
At the same time, a small increase 
(less than 10 \%) in the computational domain width results in significantly increasing flame-surface perturbations, see Fig. \ref{FimLTFU}, 
and turbulent burning velocity, see curve plotted in red dashed line in Fig. \ref{FUta}, with the increase in $U_T$ being as large as 60 \%. 
Moreover, subsequent increase in $u'$ by about 50 \% results in weakly increasing (less than 5 \%) $U_T/S_L$, see curve plotted in 
blue dotted-dashed line in Fig. \ref{FUta}.
Furthermore, normalized burning velocities obtained from turbulent flames LT/U and WT/U at large $t/\tau_f$ are close to burning velocity 
associated with the non-linear stage of development of TD instability of the three-dimensional laminar flame in the same computational domain, 
see curve plotted in yellow solid line in Fig. \ref{FlfUS}. 
These results indicate that an increase in turbulent burning velocity in cases LT/U and WT/U when compared to case LT/S 
is mainly controlled by TD instability.

Second, both flame-surface perturbations, cf. Figs. \ref{FimTFS} and \ref{FimTFU}, and dependencies of $U_T(t)/S_L$, 
cf. curves plotted in black solid and red dashed lines in Fig. \ref{FUtb}, look similar in cases T/S and T/U.
Since development of TD instability is allowed in case T/U only, these results imply that TD instability weakly affects burning velocities 
in flames T/S and T/U.
Nevertheless, $U_T/S_L$ is significantly higher in low $Le$ flame T/U when compared to counterpart equidiffusive flame T/U1, 
see curve plotted in blue dotted-dashed line in Fig. \ref{FUtb}.
This difference indicates that Lewis number substantially affects turbulent burning velocities in the discussed flames 
despite TD instability does not do so.

Third, comparison of curves plotted in black solid and blue dotted-dashed lines in Fig. \ref{FUtc} shows that differential diffusion results in 
significantly increasing $U_T(t)/S_L$ in low $Le$ flame T/UM when compared to counterpart equidiffusive flame T/UM1.
However, the effect magnitudes are comparable (i) for these two flames (a ratio of mean values of $\overline{U_T}/S_L$, obtained from flames 
TU/M and TU/M1, is about 1.9, see the most-right column in table \ref{T1}) and (ii) for two flames T/U and T/U1 (the counterpart ratio is about 
1.75) despite perturbations with the highest growth rate $\omega_{TD} \approx 5 \tau_f^{-1}$, see the most-up red circle in Fig. \ref{Flfdr}, 
are allowed in the former pair of flames only.

This interpretation is supported by comparing the ratio $R1=1.6$ of $\overline{U_T}$, obtained from equidiffusive flames T/UM1 and T/U1, 
with the counterpart ratio $R=1.7$ for low Lewis number flames T/UM and T/U.
A weak difference in $R1$ and $R$ implies that the computed increase in $U_T$ with increasing $\Lambda$ may be attributed to an increase in 
$U_T$ by the turbulence length scale $L$ to the leading order, 
whereas eventual appearance of perturbations with a high growth rate $\omega_{TD}$ plays a minor role in case T/UM.

The reported results indicate that, under the present DNS conditions, TD instability weakly affects 
turbulent burning velocity at $Ka \ge 3.4$ (cases T/UM and T/U).
However, the ratio $U_T/S_L$ is controlled by the instability at $Ka=1.6$ (case WT/U). 
These numerical findings agree well with Eq. (\ref{EKacr}), which yields $Ka_{TD}^{cr} = 1.9$ if $\tau_f \max\{\omega_{TD}(k)\} \approx 0.5$,
see the most-up red circle in Fig. \ref{Flfdr}.
Note that the critical Karlovitz number is substantially less for DL instability, because $\max\{\omega_{DL}(k)\}<\max\{\omega_{TD}(k)\}$
\citep[Fig. 2]{BGJLCAP39}.

Since development of TD instability is commonly associated with growth of flame surface area, the present results   
indicating weak contributions of TD instability to $U_T$ in flames T/U and T/UM are in line with earlier DNS \citep{BAP22b,IJHE22,CKL16} and 
experimental \citep{CFBWZHAL39} data, which showed significant (weak) influence of differential diffusion on turbulent burning velocity 
(flame surface area, respectively).
Moreover, the present results are in line with 
(i) DNS data by \citet{DBCTBL09}, which indicate that both magnitudes and length scales of fuel consumption rate perturbations differ
significantly in unstable laminar and turbulent lean hydrogen-air flames, cf. Figs. 6a and 6b in the cited paper, and
(ii) DNS data by \citet{BAP22b}, which show significantly different mean local flame characteristics in 
unstable laminar and moderately turbulent lean H$_2$-air flames, see Figs. 13, 17, 19–21, 23–25 in the cited paper.
Such DNS data imply that scales and local characteristics of unstable laminar flames are of minor value for predicting the 
counterpart scales and characteristics of turbulent flames.
Furthermore, when discussing DNS data obtained from two V-shaped weakly turbulent ($u'/S_L=0.72$ and 2.8) flames,
\citet[p. 1043]{DBBPBL09} have written that ``with increasing turbulence levels … fluctuations, 
at even the lowest intensity levels, appear to suppress to some extent the growth and propagation of the
spherical burning cells characteristic of the thermo-diffusive instability.”

Since higher values of $U_T$ obtained from low $Le$ flames T/U and T/UM when compared to their equidiffusive counterparts T/U1 and T/UM1,
respectively, do not result from TD instability, this difference in the computed burning velocities should be attributed to another physical 
mechanism, e.g., significant variations in local burning rates, caused by imbalance of molecular fluxes of chemical and thermal energies to and 
from, respectively, reactions zones stretched by turbulent eddies \citep{ZBLM,KuSa,BLL92,CRC}.
Such an effect is well pronounced in positively curved (curvature center in products) elements of flame surfaces shown in Figs. 
\ref{FimLTFU}-\ref{FimTFU} (note that the normalized local fuel consumption rate is as large as 18) and was discussed in detail by us 
recently \citep{JFM21}. 
In this regard, leading point concept \citep{KuSa,PECS05,CRC}, 
which is based on mathematically rigorous KPP theory \citep{KPP} of convection-diffusion-reaction equations, see also 
\cite{EvS00,vS03,SL15,SPL16}, appears to be the most promising approach 
to predicting the significant influence of differential diffusion on turbulent burning velocity, with such effects being documented  
both experimentally \citep{KS80,YSLWL18,CFBWZHAL39} and numerically \citep{PoF22,ADB19} even in highly turbulent flames.
Further discussion of this point is beyond the scope of this paper and the reader interested in recent advancements in 
the concept and its substantiation is referred to articles by \citet{JFM21,IJHE22} and \citet{SAL22} and references therein.

\section{Conclusions}\label{SC}

The presented DNS data indicate that the influence of thermo-diffusive instability on burning velocity is of primary (secondary) importance 
in weakly (moderately, respectively) turbulent flames. 
Under conditions of the present simulations, the instability plays an important role at Karlovitz numbers lower than a critical value, 
which is of unity order.
These DNS results agree with a simple criterion of importance of TD instability in turbulent flames, proposed recently \citep{CL23}.

\section*{Acknowledgments}
      
ANL gratefully acknowledges the financial support by Chalmers Area of Advance {\em Transport}.
Other authors have been supported in part by 
NSFC (Grant Nos. 91752201, 51976088, and 92041001), 
the Shenzhen Science and Technology Program (Grant Nos. KQTD20180411143441009 and JCYJ20210324104802005), 
Department of Science and Technology of Guangdong Province (Grant Nos. 2019B21203001 and 2020B1212030001),
National Science and Technology Major Project (Grant Nos. J2019-II-0006-0026 and J2019-II-0013-0033), 
and the Center for Computational Science and Engineering of Southern University of Science and Technology.
       
       
\bibliography{references}

\begin{thebibliography}{99}
\expandafter\ifx\csname natexlab\endcsname\relax\def\natexlab#1{#1}\fi

\bibitem{Pet32} N. Peters,
Multiscale combustion and turbulence,
Proc. Combust. Inst. {\bf 32}, 1 
 (2009).

\bibitem[Driscoll(2008)]{Dr08}
J. F. Driscoll, 
Turbulent premixed combustion: flamelet structure and its effect on turbulent burning velocities,
Prog. Energy Combust. Sci. {\bf 34}, 91 
 (2008).

\bibitem[Lipatnikov(2012)]{CRC}
A. N. Lipatnikov,
{\em Fundamentals of Premixed Turbulent Combustion}
(CRC Press, Boca Raton, FL, 2012).

\bibitem[Sabelnikov \& Lipatnikov(2017)]{ARFM} 
V. A. Sabelnikov and A. N. Lipatnikov, 
Recent advances in understanding of thermal expansion effects in premixed turbulent flames,
Annu. Rev. Fluid Mech. {\bf 49}, 91 
 (2017).

\bibitem[Driscoll et al.(2020)]{DCSCHW20}
J. F. Driscoll, J. H. Chen, A. W. Skiba, C. D. Carter, E. R. Hawkes, and H. Wang,
Premixed flames subjected to extreme turbulence: some questions and recent answers,
Prog. Energy Combust. Sci. {\bf 76}, 100802 (2017).

\bibitem{SHZ21} A. M. Steinberg, P. E. Hamlington, and X. Zhao, 
Structure and dynamics of highly turbulent premixed combustion, Prog. Energy Combust. Sci. {\bf 85}, 100900 (2021).

\bibitem[Williams(2000)]{FAW}
F. A. Williams, 
{\em Combusiton Theory}, 2nd ed. (Benjamin/Cummings, Menlo Park, CA, 1985).

\bibitem[Kuznetsov \& Sabelnikov(1990)]{KuSa}
V. R. Kuznetsov and V. A. Sabelnikov, 
{\em Turbulence and Combustion} (Hemisphere, New York, 1990).

\bibitem{JC} J. Chomiak,
{\em Combustion: a Study in Theory, Fact and Application} 
(Gordon and Breach, New York, 1990).

\bibitem[Peters(2000)]{Pet}
N. Peters, 
{\em Turbulent Combustion}
(Cambridge University Press, Cambridge, UK, 2000).

\bibitem[Zel'dovich et~al.(1985)]{ZBLM}
Ya.~B. Zel'dovich, G.~I. Barenblatt, V.~B. Librovich, and G.~M. Makhviladze, 
\textit{The Mathematical Theory of Combustion and Explosions}
(Consultants Bureau, New Tork, 1985).

\bibitem[Landau \& Lifshitz(1987)]{LL}  
L. D. Landau and E.~M. Lifshitz,
{\em Fluid Mechanics} (Pergamon Press, Oxford, 1987).

\bibitem[Clavin(1985)]{Cl85}
P. Clavin, 
Dynamical behavior of premixed flame fronts in laminar and turbulent flows,
Prog. Energy Combust. Sci. {\bf 11}, 1 
 (1985).

\bibitem{KH05} 
S. Kadowaki and T. Hasegawa,
Numerical simulation of dynamics of premixed flames: flame instability and vortex-flame interaction,
Prog. Energy Combust. Sci. {\bf 31}, 193 
 (2005).

\bibitem[Matalon(2007)]{Mat07}
M. Matalon,
Intrinsic flame instabilities in premixed and nonpremixed combustion,
Annu. Rev. Fluid Mech. {\bf 39}, 163 
 (2007).

\bibitem[Hochgreb(2023)]{H39}
S. Hochgreb, 
How fast can we burn, 2.0.
Proc. Combust. Inst. {\bf 39}, in press.

\bibitem[Karpov \& Sokolik(1961)]{KS61}
V.~P. Karpov and A.~S. Sokolik,
Ignition limits in turbulent gas mixtures,
Proc. Acad. Sci. USSR, Phys. Chem. {\bf 141}, 866 
 (1961).

\bibitem[Karpov \& Severin(1980)]{KS80}
V.~P. Karpov and E.~S. Severin, 
Effects of molecular-transport coefficients on the rate of turbulent combustion,
Combust. Explos. Shock Waves {\bf 16}, 41 
 (1980).

\bibitem[Lipatnikov \& Chomiak(2005)]{PECS05}
A. N. Lipatnikov and J. Chomiak, 
Molecular transport effects on turbulent flame propagation and structure,
Prog. Energy Combust. Sci. {\bf 31}, 1 
 (2005).

\bibitem[Yang et al.(2018)]{YSLWL18}
S. Yang, A. Saha, W. Liang, F.  Wu, and C. K. Law,
Extreme role of preferential diffusion in turbulent flame propagation,
Combust. Flame {\bf 188}, 498 
 (2018).

\bibitem[Nguyen et al.(2019)]{NYS37}
M. T. Nguyen, D. W. Yu, and S. S. Shy,
General correlations of high pressure turbulent burning velocities with the consideration of Lewis number effect,
Proc. Combust. Inst. {\bf 37}, 2391 
 (2019).

\bibitem{LCS39} 
A. N. Lipatnikov, Y.-R. Chen, and S. S. Shy, 
An experimental study of the influence of Lewis number on turbulent flame speed at different pressures, 
Proc. Combust. Inst. {\bf 39}, in press.

\bibitem[Venkateswaran et al.(2011)]{VMSNSL11}
P. Venkateswaran, A. Marshall, D.~H. Shin, D. Noble, J. Seitzman, and T. Lieuwen,
Measurements and analysis of turbulent consumption speeds of H$_2$/CO mixtures,
Combust. Flame {\bf 158}, 1602 
 (2011).

\bibitem[Daniele et al.(2011)]{DJMB33}
S. Daniele, P. Jansohn, J. Mantzaras, and K. Boulouchos,
Turbulent flame speed for syngas at gas turbine relevant conditions, 
Proc. Combust. Inst. {\bf 33}, 2937 
 (2011).

\bibitem[Venkateswaran et al.(2013)]{VMSL34}
P. Venkateswaran, A. Marshall, J. Seitzman, and T. Lieuwen,
Pressure and fuel effects on turbulent consumption speeds of H$_2$/CO blends,
Proc. Combust. Inst. {\bf 34}, 1527 
 (2013).

\bibitem[Zhang et al.(2018)]{ZWYJZH18}  
W. Zhang, J. Wang, Q. Yu, W. Jin, M. Zhang, and Z. Huang, 
Investigation of the fuel effects on burning velocity and flame structure of turbulent premixed flames based on leading points concept,
Combust. Sci. Technol. {\bf 190}, 1354 
 (2018).

\bibitem[Xia et al.(2020)]{XHHHHKF20}
Y. Xia, G. Hashimoto, K. Hadi, N. Hashimoto, A. Hayakawa, H. Kobayashi, and O. Fujita,
Turbulent burning velocity of ammonia/oxygen/nitrogen premixed flame in O$_2$-enriched air condition,
Fuel {\bf 268}, 117383 (2020).

\bibitem{LBCMR38} C. Lhuillier, P. Brequigny, F. Contino, and C. Mounaim-Rousselle, 
Experimental investigation on ammonia combustion behavior in a spark-ignition engine by means of laminar and turbulent expanding flames,
Proc. Combust. Inst. {\bf 38}, 5859 (2021).

\bibitem[Cai et al.(2022)]{CFBWZHAL39}
X. Cai, Q. Fan, X.-S. Bai, J. Wang, M. Zhang, Z. Huang, M. Ald\'en, and Z. Li,
Turbulent burning velocity and its related statistics of ammonia-hydrogen-air jet flames at high Karlovitz number: 
Effect of differential diffusion,
Proc. Combust. Inst. {\bf 39}, in press.

\bibitem{WEAR23}
S. Wang, A. M. Elbaz, O. Z. Arab, and W. L. Roberts, 
Turbulent flame speed measurement of NH$_3$/H$_2$/air and CH$_4$/air flames and a numerical case study of NO emission in 
a constant volume combustion chamber, 
Fuel {\bf 332}, 126152 (2023).
 
\bibitem{WEWWR23}
S. Wang, A. M. Elbaz, G. Wang, Z. Wang, and W. L. Roberts, 
Turbulent flame speed of NH$_3$/CH$_4$/H$_2$/H$_2$O/air-mixtures: Effects of elevated pressure and Lewis number, 
Combust. Flame {\bf 247}, 112488 (2023).

\bibitem[Aspden et al.(2011)]{ADB11}
A. J. Aspden, M.~S. Day, and J.~B. Bell,
Turbulence-flame interactions in lean premixed hydrogen: transition to the distributed burning regime,
J. Fluid Mech. {\bf 680}, 287 
 (2011).

\bibitem[Chakraborty \& Cant(2011)]{CC11}
N. Chakraborty and R. S. Cant,
Effects of Lewis number on flame surface density transport in turbulent premixed combustion,
Combust. Flame {\bf 158}, 1768 
 (2011).

\bibitem[Kobayashi et al.(2013)]{KOWMOOKK34}
H. Kobayashi, Y. Otawara, J. Wang, F. Matsuno, Y. Ogami, M. Okuyama, T. Kudo, and S. Kadowaki,
Turbulent premixed flame characteristics of a CO/H$_2$/O$_2$ mixture highly diluted with CO$_2$ in a high-pressure environment,
Proc. Combust. Inst. {\bf 36}, 1437 
 (2013).

\bibitem[Dopazo et al.(2017)]{DCC17}
C. Dopazo, L. Cifuentes, and N. Chakraborty, 
Vorticity budgets in premixed combusting turbulent flows at different Lewis numbers,
Phys. Fluids {\bf 29}, 045106 (2017).

\bibitem[Zhang et al.(2020)]{ZWH20}
M. Zhang, J. Wang, and Z. Huang, 
Turbulent flame structure characteristics of hydrogen enriched natural gas with CO$_2$ dilution,
Int. J. Hydrogen Energy {\bf 45}, 20426 
 (2020).

\bibitem[Berger et al.(2022b)]{BAP22b} 
L. Berger, A. Attili, and H. Pitsch,
Synergistic interactions of thermodiffusive instabilities and turbulence in lean hydrogen flames,
Combust. Flame {\bf 244}, 112254 (2022).

\bibitem[Bradley et al.(1992)]{BLL92}
D. Bradley, A.~K.~C. Lau, and M. Lawes, 
Flame stretch rate as a determinant of turbulent burning velocity,
Phil. Trans. R. Soc. London A {\bf 338}, 359 
 (1992).

\bibitem{CL23}
J. Chomiak and A. N. Lipatnikov,
A simple criterion of importance of thermo-diffusive instability in premixed turbulent flames,
Phys. Rev. E {\bf 107}, 015102 (2023).

\bibitem[Buckmaster \& Ludford(1982)]{BL82}
J.~D. Buckmaster and G.~S.~S. Ludford, 
{\em Theory of Laminar Flames}
(Cambridge Univ. Press, Cambridge, UK, 1982).

\bibitem[Sivashinsky et al.(1982)]{SLJ82}
G.~I. Sivashinsky, C.~K. Law, and G. Joulin, 
On stability of premixed flames in stagnation-point flow,
Combust. Sci. Technol. {\bf 28}, 155 
(1982).

\bibitem[Korsarts et al.(1997)]{KBS97}
Y. Korsarts, J. Brailovsky, and G.~I. Sivashinsky, 
On hydrodynamic instability of stretched flames,
Combust. Sci. Technol. {\bf 123}, 207 
(1997).

\bibitem[Altantzis et al.(2011)]{AFTKB33}
C. Altantzis, C. E. Frouzakis, A.~G. Tomboulides, K. Kerkemeier, and K. Boulouchos,
Detailed numerical simulations of intrinsically unstable two-dimensional planar lean premixed hydrogen/air flames,
Proc. Combust. Inst. {\bf 33}, 1261 
 (2011).

\bibitem[Altantzis et al.(2012)]{AFTMB12}
C. Altantzis, C. E. Frouzakis, A.~G. Tomboulides, M. Matalon, and K. Boulouchos, 
Hydrodynamic and thermodiffusive instability effects on the evolution of laminar planar lean premixed hydrogen flames,
J. Fluid Mech. {\bf 700}, 329 
 (2012).

\bibitem[Frouzakis et al.(2015)]{FFTAM35}
C. E. Frouzakis, N. Fogla, A.~G. Tomboulides, C. Altantzis, and M. Matalon, 
Numerical study of unstable hydrogen/air flames: shape and propagation speed,
Proc. Combust. Inst. {\bf 35}, 1087 
 (2015).

\bibitem[Berger et al.(2019)]{BKAP19}
L. Berger, K. Kleinheinz, A. Attili, and H. Pitsch,
Characteristic patterns of thermodiffusively unstable premixed lean hydrogen flames,
Proc. Combust. Inst. {\bf 37}, 1879 
 (2019).

\bibitem[Berger et al.(2022a)]{BAP22a} 
L. Berger, A. Attili, and H. Pitsch,
Intrinsic instabilities in premixed hydrogen flames: Parametric variation of pressure, equivalence ratio, and temperature,
Part 1 - Dispersion relations in the linear regime,
Combust. Flame {\bf 240}, 111935 (2022).

\bibitem[Berger et al.(2023)]{BGJLCAP39}
L. Berger, M. Grinberg, B. J\"urgens, P.~E. Lapenna, F. Creta, A. Attili, and H. Pitsch,       
Flame fingers and interactions of hydrodynamic and thermodiffusive instabilities in laminar lean hydrogen flames,
Proc. Combust. Inst. {\bf 39}, in press.

\bibitem{MY} 
A. S. Monin and A. M. Yaglom, 
{\em Statistical Fluid Mechanics: Mechanics of Turbulence}, vol. 2
(The MIT Press, Cambridge, MA, 1975).

\bibitem{Frisch}
U. Frisch, {\em Turbulence. The Legacy of A.N. Kolmogorov} 
(Cambridge University Press, Cambridge, UK, 1995).

\bibitem[Boughanem \& Trouv\'e(1998)]{BT27}
H. Boughanem and A. Trouv\'e, 
The domain of influence of flame instabilities in turbulent premixed combustion,
Proc. Combust. Inst. {\bf 27}, 971 
 (1998).

\bibitem[Chaudhuri et al.(2011)]{CAL11}
S. Chaudhuri, V. Akkerman, and C. K. Law,
Spectral formulation of turbulent flame speed with consideration of hydrodynamic instability,
Phys. Rev. E {\bf 84}, 026322 (2011). 

\bibitem[Creta \& Matalon(2011)]{CM11}
F. Creta and M. Matalon, 
Propagation of wrinkled turbulent flames in the context of hydrodynamic theory,
J. Fluid Mech. {\bf 680}, 225 
 (2011).

\bibitem[Fogla et~al.(2013)]{FCM34}
N. Fogla, F. Creta, and M. Matalon,
Influence of the Darrieus-Landau instability on the propagation of planar turbulent flames,
Proc. Combust. Inst. {\bf 34}, 1509 
 (2013).

\bibitem[Fogla et~al.(2015)]{FCM15}
N. Fogla, F. Creta, and M. Matalon,      
Effect of folds and pockets on the topology and propagation of premixed turbulent flames,
Combust. Flame {\bf 162}, 2758 
 (2015).

\bibitem[Fogla et~al.(2017)]{FCM17}
N. Fogla, F. Creta, and M. Matalon, 
The turbulent flame speed for low-to-moderate turbulence intensities: Hydrodynamic theory vs. experiments,
Combust. Flame {\bf 175}, 155 
 (2017).

\bibitem{LTLC38}
P. E. Lapenna, G. Troiani, R. Lamioni, and F. Creta,
Mitigation of Darrieus–Landau instability effects on turbulent premixed flames,
Proc. Combust. Inst. {\bf 38}, 2885 
 (2021).

\bibitem[Pelc\'e \& Clavin(1982)]{PC82}
P. Pelc\'e and P.Clavin,
Influence of hydrodynamics and diffusion upon the stability limits of laminar premixed flames,
J. Fluid Mech. {\bf 124}, 219 
 (1982).

\bibitem[Matalon \& Matkowsky(1982)]{MM82}
M. Matalon and B.~J. Matkowsky, 
Flames as gas dynamic discontinuities,
J. Fluid Mech. {\bf 124}, 239 
 (1982).

\bibitem[Frankel \& Sivashinsky(1982)]{FS82}
M. L. Frankel and G. J. Sivashinsky,
The effect of viscosity on hydrodynamic stability of a plane flame front,
Combust. Sci. Technol. {\bf 29}, 207 
 (1982).

\bibitem[Class et al.(2003)]{CMK03}
A.~G. Class, B.~J. Matkowsky, and A.~Y. Klimenko, 
Stability of planar flames as gasdynamic discontinuities,
J. Fluid Mech. {\bf 491}, 51 
 (2003).

\bibitem[Kelley et al.(2012)]{KBL12}        
A.~P. Kelley, J.~K. Bechtold, and C.~K. Law, 
Premixed flame propagation in a confining vessel with weak pressure rise,
J. Fluid Mech. {\bf 691}, 26 
 (2012).

\bibitem[Clanet \& Searby(1998)]{CS98}
C. Clanet and G. Searby, 
First experimental study of the Darrieus-Landau instability,
Phys. Rev. Let. {\bf 80}, 3867 
 (1998).

\bibitem[Truffaut \& Searby(1999)]{TS99}
J.-M. Truffaut and G. Searby, 
Experimental study of the Darrieus-Landau instability on an inverted-V flame, and measurement of the Markstein number,
Combust. Sci. Technol. {\bf 149}, 35 
 (1999).

\bibitem{YBB15}
R. Yu, X.-S. Bai, and V. Bychkov, 
Fractal flame structure due to the hydrodynamic Darrieus-Landau instability,
Phys. Rev. E {\bf 92}, 063028 (2015).

\bibitem{CLLFM20} F. Creta, P. E. Lapenna, R. Lamioni, N. Fogla, and M. Matalon,
Propagation of premixed flames in the presence of Darrieus–Landau and thermal diffusive instabilities,
Combust. Flame {\bf 216}, 256 
 (2020).

\bibitem[Berger et al.(2022c)]{BAP22c}
L. Berger, A. Attili, and H. Pitsch,
Intrinsic instabilities in premixed hydrogen flames: Parametric variation of pressure, equivalence ratio, and temperature,
Part 2 - Non-linear regime and flame speed enhancement,
Combust. Flame {\bf 240}, 111936 (2022).

\bibitem[Lee et al.(2021)]{JFM21}
H.~C. Lee, P. Dai, M. Wan, and A. N. Lipatnikov, 
Influence of molecular transport on burning rate and conditioned species concentrations in highly turbulent premixed flames,
J. Fluid Mech. {\bf 298}, A5 (2021).

\bibitem[Lee et al.(2022a)]{CNF22PI}
H.~C. Lee, P. Dai, M. Wan, and A. N. Lipatnikov,
A DNS study of extreme and leading points in lean hydrogen-air turbulent flames - part I:
Local thermochemical structure and reaction rates, 
Combust. Flame {\bf 235}, 111716 (2022). 

\bibitem[Lee et al.(2022b)]{CNF22PII} 
H.~C. Lee, P. Dai, M. Wan, and A. N. Lipatnikov,         
A DNS study of extreme and leading points in lean hydrogen-air turbulent flames - part II: 
Local velocity field and flame topology, 
Combust. Flame {\bf 235}, 111712 (2022).         
 
\bibitem[Lee et al.(2022c)]{PoF22}
H.~C. Lee, P. Dai, M. Wan, and A. N. Lipatnikov,        
Lewis number and preferential diffusion effects in lean hydrogen–air highly turbulent flames,
Phys. Fluids {\bf 34}, 035131 (2022).

\bibitem[Lee et al.(2022d)]{IJHE22}
H.~C. Lee, P. Dai, M. Wan, and A. N. Lipatnikov,         
A numerical support of leading point concept,
Int. J. Hydrogen Energy {\bf 47}, 23444 
 (2022).

\bibitem[Lee et al.(2022e)]{Fuel22}
H.~C. Lee, A. Abdelsamie, P. Dai, M. Wan, and A. N. Lipatnikov,
Influence of equivalence ratio on turbulent burning velocity and extreme fuel consumption rate in lean hydrogen-air turbulent flames,
Fuel {\bf 327}, 124969 (2022).

\bibitem[K\'eromn\`es et al.(2013)]{Curran}
A. K\'eromn\`es, W.~K. Metcalfe, K.~A. Heufer, N. Donohoe, A.~K. Das, C.-J. Sung, J. Herzler, C. Naumann, P. Griebel, O. Mathieu, 
M.~C. Krejci, E.~L. Petersen, W.~J. Pitz, and H.~J. Curran, 
An experimental and detailed chemical kinetic modeling study of hydrogen and syngas mixture oxidation at elevated pressures,
Combust. Flame {\bf 160}, 995 
(2013).

\bibitem[Goodwin et al.(2009)]{Cantera}
D. Goodwin, N. Malaya, H. Moffat, and R. Speth,
Cantera: An object-oriented software toolkit for chemical kinetics, thermodynamics, and transport processes
(Caltech, Pasadena, CA, 2009).

\bibitem[Abdelsamie et al.(2016)]{DINO}
A. Abdelsamie, G. Fru, T. Oster, F. Dietzsch, G. Janiga, and D. Th\'evenin,
Towards direct numerical simulations of low-Mach number turbulent reacting and two-phase flows using immersed boundaries,
Comput. Fluids {\bf 131}, 123 
(2016).

\bibitem[Lundgren(2003)]{Lu03}
T. Lundgren, 
Linearly forced isotropic turbulence
(Tech. Rep., Minnesota University of Minneapolis, 2003).

\bibitem[Rosales \& Meneveau(2005)]{RM05}
C. Rosales and C. Meneveau,
Linear forcing in numerical simulations of isotropic turbulence: Physical space implementations and convergence properties,
Phys. Fluids {\bf 17}, 095106 (2005).

\bibitem[Carroll \& Blanquart(2014)]{CB14}
P.~L. Carroll and G. Blanquart, 
The effect of velocity field forcing techniques on the Karman–Howarth equation,
J. Turbul. {\bf 15}, 429 
(2014).

\bibitem[Lipatnikov \& Chomiak(2002)]{PECS02}
A.~N. Lipatnikov and J. Chomiak, 
Turbulent flame speed and thickness: phenomenology, evaluation, and application in multi-dimensional simulations,
Prog. Energy Combust. Sci. {\bf 28}, 1 
(2002).

\bibitem[Yu \& Lipatnikov(2017)]{YL17}
R. Yu and A.~N. Lipatnikov, 
DNS study of dependence of bulk consumption velocity in a constant-density reacting flow on turbulence and mixture characteristics,
Phys. Fluids {\bf 29}, 065116 (2017).

\bibitem[Kim et al.(2020)]{KSLG20} J. Kim, A. Satija, R. P. Lucht, and J. P. Gore,
Effects of turbulent flow regime on perforated plate stabilized piloted lean premixed flames,
Combust. Flame {\bf 211}, 158 (2020).

\bibitem[Chakraborty et al.(2016)]{CKL16}
N. Chakraborty, I. Konstantinou, and A.N. Lipatnikov,
Effects of Lewis number on vorticity and enstrophy transport in turbulent premixed flames,
Phys. Fluids {\bf 28}, 015109 (2016).

\bibitem[Day et al.(2009a)]{DBCTBL09} M. S. Day, J. B. Bell, R. K. Cheng, S. Tachibana, V. E. Beckner, and M. J. Lijewski, 
Cellular burning in lean premixed turbulent hydrogen-air flames: coupling experimental and computational analysis at the laboratory scale.
J. Phys.: Conf. Ser. {\bf 180}, 012031 (2009).

\bibitem[Day et al.(2009b)]{DBBPBL09} M. S. Day, J. B. Bell, P. T. Bremer, V. Pascucci, V. E. Beckner, and M. J. Lijewski, 
Turbulence effects on cellular burning structures in lean premixed hydrogen flames,
Combust. Flames {\bf 156}, 1035 (2009).

\bibitem[Kolmogorov et al.(1937)]{KPP}
A.~N. Kolmogorov, E.~G. Petrovsky, and N.~S. Piskounov,  
A study of the diffusion equation with a source term and its application to a biological problem,
Bjul. MGU Section A {\bf 1(6)}, 1 
(1937).

\bibitem[Ebert \& van Saarloos(2000)]{EvS00}
U. Ebert and W. van Saarlos, 
Front propagation into unstable states: universal algebraic convergence towards uniformly translating pulled fronts,
Physica D {\bf 146} 1 
(2000).

\bibitem[van Saarloos(2003)]{vS03}
W. van Saarloos, 
Front propagation into unstable states,
Phys. Reports {\bf 386}, 29 
(2003).

\bibitem[Sabelnikov \& Lipatnikov(2015)]{SL15}
V.~A. Sabelnikov and A.~N. Lipatnikov, 
Transition from pulled to pushed fronts in premixed turbulent combustion: theoretical and numerical study,
Combust. Flame {\bf 162}, 2893 
(2015).

\bibitem[Sabelnikov et al.(2016)]{SPL16}
V.~A. Sabelnikov, N.~N. Petrova, and A.~N. Lipatnikov,
Analytical and numerical study of travelling waves using the Maxwell-Cattaneo relaxation model extended to reaction-advection-diffusion systems,
Phys. Rev. E {\bf 94}, 042218 (2016).

\bibitem[Aspden et al.(2019)]{ADB19}
A.~J. Aspden, M.~S. Day, and J.~B. Bell, 
Towards the distributed burning regime in turbulent premixed flames,
J. Fluid Mech. {\bf 871}, 1 
(2019).

\bibitem[Somappa et al.(2022)]{SAL22}
S. Somappa, V. Acharia, and T. Lieuwen,
Finite flame thickness effects on KPP turbulent burning velocities,
Phys. Rev. E {\bf 106}, 055107 (2022).

%
%
\end{thebibliography}
  
\end{document}